\documentclass[conference]{IEEEtran}
\usepackage[english]{babel}
\usepackage[algo2e,ruled,linesnumbered]{algorithm2e}
\usepackage{import}
\usepackage[]{units}
\usepackage{url}
\usepackage{subfigure}
\usepackage{textcomp}
\SetKwComment{Comment}{$\triangleright$\ }{}
\ifCLASSINFOpdf
  \usepackage[pdftex]{graphicx}
\else
\fi
\usepackage[cmex10]{amsmath}

\usepackage[absolute,showboxes]{textpos}

\TPMargin{5pt}

\newcommand{\copyrightstatement}{
    \begin{textblock}{0.84}(0.08,0.01)    
         \noindent
         \footnotesize
         \copyright 2019 IEEE. Personal use of this material is permitted. Permission from IEEE must be obtained for all other uses, in any current or future media, including reprinting/republishing this material for advertising or promotional purposes, creating new collective works, for resale or redistribution to servers or lists, or reuse of any copyrighted component of this work in other works.
    \end{textblock}
}

\usepackage[inline]{enumitem}

\usepackage[np]{numprint}
\npstyleenglish

\usepackage{amssymb}
\usepackage{xcolor}

\graphicspath{{figures/}}
\begin{document}
	
\bstctlcite{IEEEexample:BSTcontrol}

\copyrightstatement

%
\title{HNLB: Utilizing Hardware Matching Capabilities of NICs for Offloading Stateful Load Balancers}
\author{

\IEEEauthorblockN{Raphael Durner, Amir Varasteh, Max Stephan, Carmen Mas Machuca, and Wolfgang Kellerer}
\IEEEauthorblockA{Chair of Communication Networks\\
Department of Electrical and Computer Engineering\\
Technical University of Munich, Germany\\
Email: \{r.durner, amir.varasteh, maximilian.stephan, cmas, wolfgang.kellerer\}@tum.de}

}

\maketitle

\begin{abstract}

In order to scale web  or other services, the load on single instances of the respective service has to be balanced.
Many services are stateful such that packets belonging to the same connection must be delivered to the same instance.
This requires stateful load balancers which are mostly implemented in software.
On the one hand, modern packet processing frameworks supporting software load balancers, such as the Data Plane Development Kit (DPDK), deliver  high performance compared  to older approaches.
On the other hand, common Network Interface Cards (NICs) provide additional matching capabilities that can be utilized for increasing the performance even further and in turn reduce the necessary server resources. In fact, offloading the packet matching to hardware can free up CPU cycles of the servers.

Therefore, in this work, we propose the Hybrid NIC-offloading Load Balancer (HNLB), a high performance hybrid hardware-software load balancer, utilizing the NIC-offloading hardware matching capabilities.
The results of our performance evaluations show that the throughput using NIC offloading can be increased by up to 50\%, compared to a high performance software-only implementation.
Furthermore, we investigated the limitations of our proposed approach, e.g., the limited number of possible concurrent connections. 
\end{abstract}


%


\section{Introduction} \label{sec:introduction} 

Today, with advances in Internet and Cloud Computing, Data Center (DC) traffic volume is growing exponentially. For instance, it is reported that Google DCs traffic has been increased fifty times from 2008 to 2014 \cite{singh2015jupiter}. There are hundreds to thousands of servers in a DC that are providing an ever increasing array of services \cite{roy2015inside,singh2015jupiter}. According to \cite{patel2013ananta}, 44$\%$ of the cloud traffic needs load balancing. 
This has made load balancing one of the most important network services in a cloud DC. 

Traditional load balancers were deployed via dedicated hardware \cite{a10,f5,lb} in the network.
With increasing network demands it got more and more difficult to serve them due to a lack in flexibility of the hardware functions.
As a relief load balancers have been started to be deployed in software.
Besides improving manageability and flexibility, Software Load Balancers (SLB) can be easily deployed in modern data center architectures \cite{eisenbud2016maglev}.
However, SLBs have two fundamental limitations: high resource consumption, and high latency \cite{gandhi2015duet}.

To overcome these limitations, we propose the Hybrid NIC-offloading Load Balancer (HNLB), an efficient hybrid hardware-software load balancer which aims to flexibly combine hardware with software packet processing.
HNLB utilizes NIC offloading and seamlessly combines it with software implementations as described in this paper. 
In addition, HNLB does not require specialized hardware, but only utilizes NICs already available on many servers today. 

NIC offloading has not been receiving much attention, although it can bring several advantages in different ways. 
For instance, using the NIC's on-board computation power, the load on the CPU can be reduced significantly, which leads to an increase in the capacity of the solution. 
Additionally, we expect lower latency since the processing of one packet is done in fewer processor cycles.
We argue that a smart combination of the advantages of software (high flexibility) and hardware (high performance) can bring us closer to solving the challenges of next generation data centers serving 100G+ per server. 

\begin{figure}[t]
	\centering
	\includegraphics[width=0.6\columnwidth]{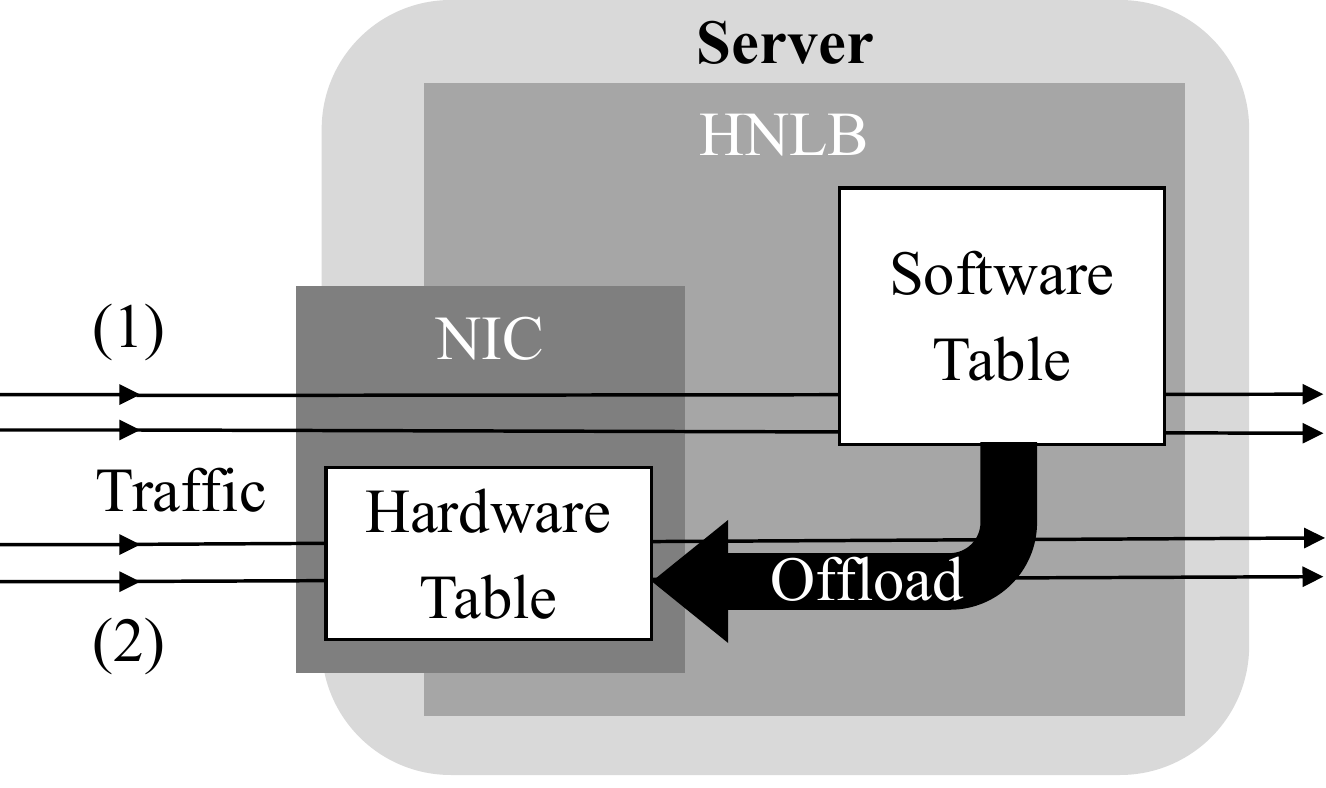}
	\vspace{-0.3cm}
	\caption{HNLB utilizes NICs for offloading packet matching capabilities to hardware.}
	\vspace{-0.5cm}
	\label{fig:idea}
\end{figure}

The core idea of HNLB is presented in Figure \ref{fig:idea}. Every stateful load balancer has to keep a table storing the state of connections.
An SLB matches incoming packets against its Software Table and forwards the packet to the endpoint responsible for this connection.
We utilize the Hardware Table of NICs by offloading the Software Table to this Hardware Table.
This aims to reduce load on the CPU, since the packet matching is offloaded to fast hardware.
Note that the NIC cannot fully takeover the processing of packets, since it is not able to perform packet forwarding.
Our evaluation results indicate that a reduction of the load on the CPU can increase the maximum achievable throughput significantly.


More specifically, we propose to utilize Flow Director NIC technology~\cite{flowdirector2014intel} to offload packet matching task of the load balancer to server's NIC. 
This technology is available on many modern Intel NIC chip-sets and as well from other manufacturers~\cite{mex}.
Hence, we argue that the hardware is already available today in many existing DCs. However, their capabilities are not utilized to their full extent.

Further, other stateful packet processing software, such as NATs, stateful Firewalls or accounting functions need Software Tables. Hence, they can gain performance in the same way from NIC offloading.

In this work, we implement, measure, and evaluate the performance gains of NIC offloading in load balancer network service.
Therefore, the main contributions of this work are:
\begin{itemize}
	\item[1)] We propose HNLB, a high performance load balancer utilizing NIC offloading
	\item[2)] The gain of NIC offloading considering different scenarios is investigated through comparison with an SLB.
	\item[3)] Finally we present a novel utilization metric to be able to measure the load of both approaches. 
\end{itemize}


The rest of this paper is organized as follows: Section \ref{sec:background} provides background on load balancing and Intel Flow Director. In Section \ref{sec:Arch}, our proposed system architecture and testbed is introduced. We propose a novel utilization metric in Section \ref{sec:ass}. Section \ref{sec:eval} presents the performance evaluation of HNLB. In Section \ref{sec:dissc}, we discuss and analyze the results, followed by a review of related work in Section \ref{sec:related_work}. Finally, we draw a conclusion and present our planned next steps in Section \ref{sec:conc}.

\section{Background} \label{sec:background} 

In this section we provide background on stateful load balancers and introduce briefly Intel's FlowDirector technology .

\subsection{Stateful Load Balancers}
The main goal of a load balancer is to distribute the load (i.e., the packets) between several back-end servers (service instances) that deliver the actual service.
The packets can be either distributed stateless (e.g., using round robin) or stateful (packets belonging to the same connection are always delivered to the same back-end server).
Further, existing load balancers work on different layers, e.g. layer~4 (L4) and layer~7.

In this work, we utilize connection table offloading on NICs that support header matching. 
Thus, we focus on the design and evaluation of a stateful L4 load balancer.

Two different sets of IPs exist in the load balancer concept: \textit{i)} Virtual IPs (VIPs), and \textit{ii)} Direct IPs (DIPs). 
VIPs are the IPs that the users are addressing. 
They can be seen as the service addresses.
On the other hand, DIPs are the addresses that belong to the actual instances delivering the service.
Consequently, the load balancer distributes the packets destined to one VIP between several DIPs.

Accordingly, in order to guarantee connection consistency, a stateful L4 load balancer has to maintain two tables:

\begin{itemize}
	\item[I)] The VIP table: which contains the VIPs of all services and the respective active DIPs belonging to this service.
	\item[II)] The connection table: which holds the mapping of active connections to their designated DIPs.
\end{itemize}

\begin{figure}[t]
    \centering
	\includegraphics[width=1\columnwidth]{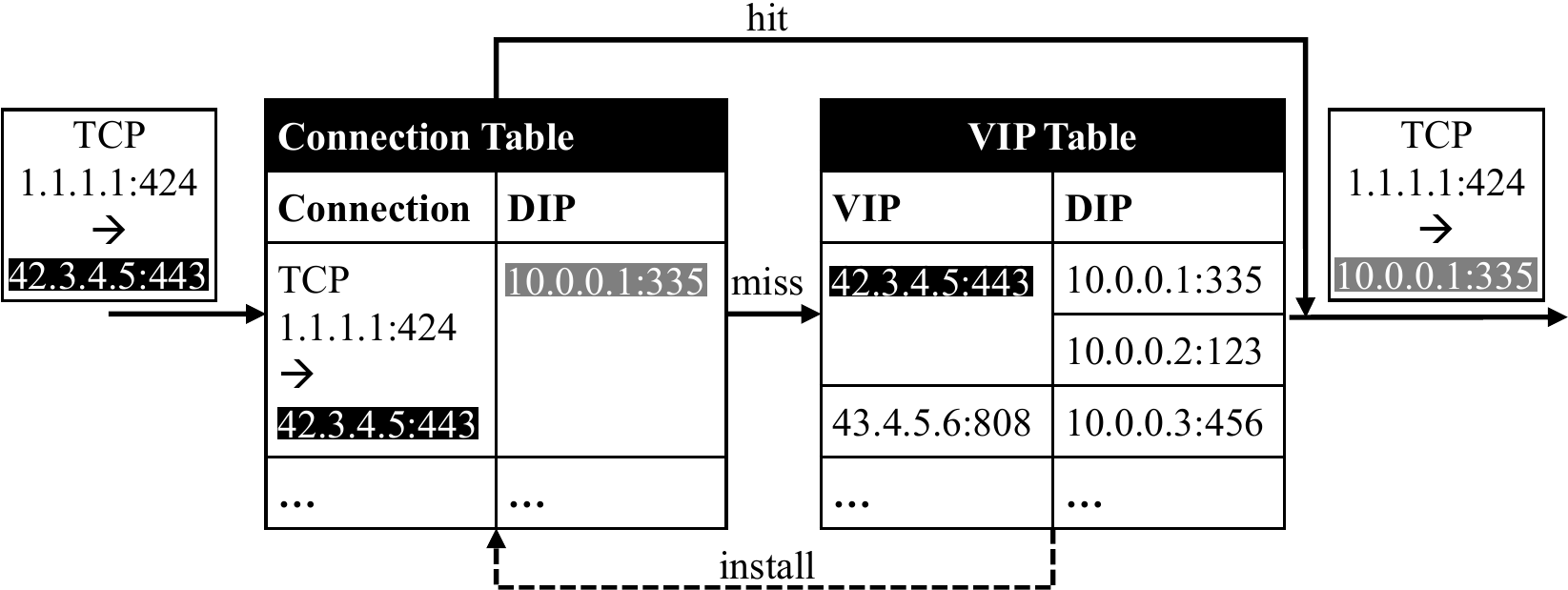}
	\vspace{-0.3cm}
	\caption{Basic working principle of a stateful L4 load balancer. The VIP of the served packet is highlighted in black, the DIP is highlighted in gray. }
	\label{fig:LB}
	\vspace{-0.3cm}
\end{figure}

The basic working principle of a stateful L4 load balancer is shown in Figure \ref{fig:LB}:
Incoming packets are first checked against the Connection Table. If an entry already exists (\textit{hit}), the packet header is rewritten according to the entry, and the packet is forwarded.
If the 5-tuple of the packet does not match any entry (\textit{miss}) in the Connection Table, a new DIP for the corresponding VIP is selected from the VIP table. 
This selection can be based on round robin algorithm or according to the current load of the DIPs.
Afterwards, the packet is rewritten using the chosen mapping and forwarded.
Finally, the new connection is installed in the Connection Table (\textit{install}).

The figure shows an incoming packet with the 5-tuple: TCP,~1.1.1.1,~424,~42.3.4.5,~443, that directly matches an entry in the Connection Table (hit).
Thus the packet's destination is rewritten to the DIP 10.0.0.1 with destination port 335.

\subsection{Intel Flow Director}

In order to support multi-core packet processing, packets have to be distributed among the cores.
One technique that can achieve this is called Receive Side Scaling (RSS).
RSS firstly computes the hash of the packet header. 
It then forwards the packets according to the hash value to one of the NIC queues.
Each core is then processing packets of one queue.
This procedure constitutes a stateless load balancer.
The Flow Director technology is originally designed to expand the RSS functionality, by not only performing load balancing, but also forwarding the incoming packets to the core where the related application is running. 
Additionally, the match table can be programmed using an API. 
In HNLB, we utilize this functionality to offload the connection table of the load balancer to the NIC.

Depending on the configuration of the NIC, Flow Director can support between around 2000 and 8000 table entries \cite{intelx540}. 
As a drawback the NIC's memory is shared between the Flow Director and the receive buffer.
Therefore, as the number rules is increased, the receive buffer is shrinked.

We measured the latency of a Flow Director enabled echo software for different number of filters. 
The absolute maximum latency that occurred in our tests increased from $95\,\mu s$ for no rules to $105\,\mu s$ for 8000 rules.
These results clearly show that the induced latency by Flow Director is marginal, even for 8000 rules.

\section{Implementation}
\label{sec:Arch}

In this section we present how HNLB and the SLB is implemented in detail.

\begin{figure}[t]
    \centering
	\includegraphics[width=0.95\columnwidth]{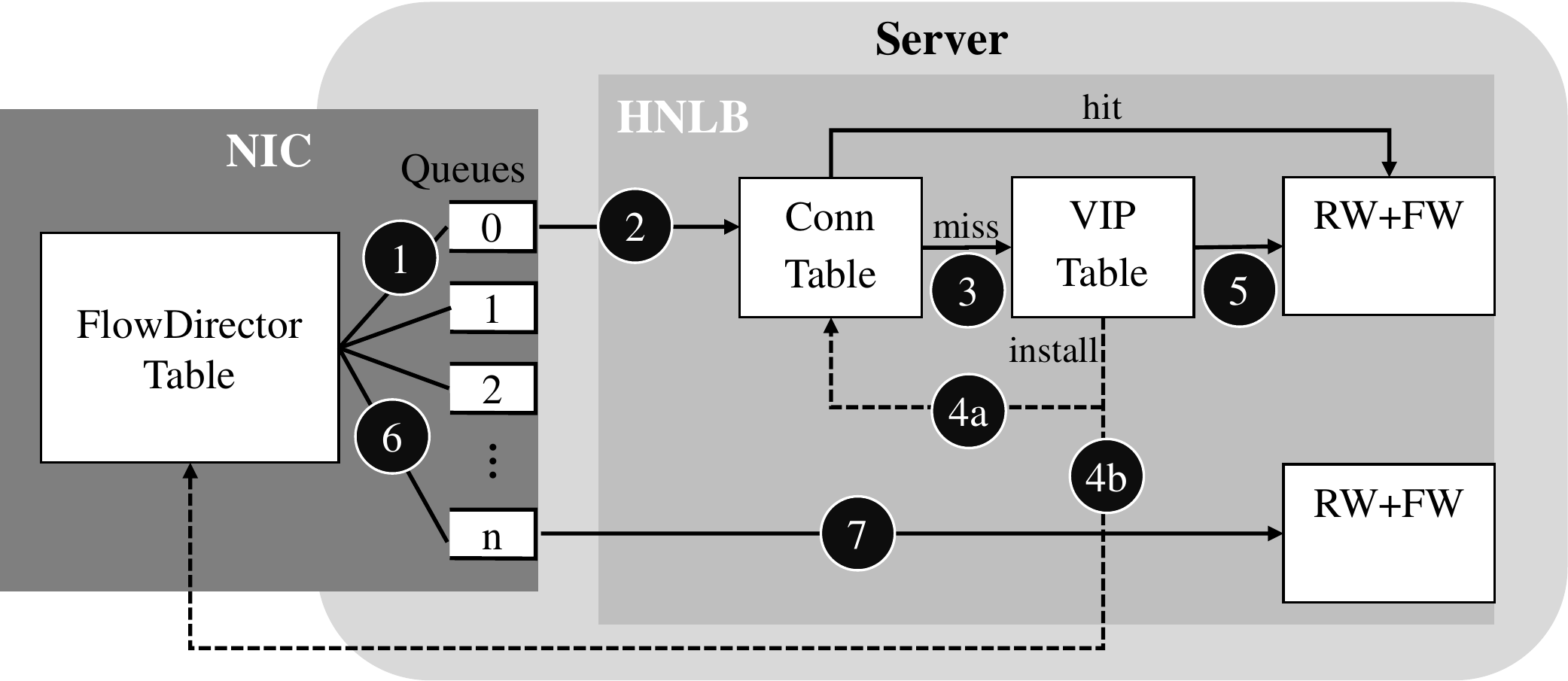}
	\caption{Load balancer with hardware table offloading: Packets from unseen connections are put into queue 0 (1) and polled from the HNLB Software (2). These packets do not match any connection in the software connection table (3), therefore a DIP is chosen from the VIP table and the mapping is installed in software (4a) and hardware (4b). The packet header is rewritten and forwarded (5). Subsequent packets of this connection are put in the queue of the corresponding DIP (6). When they are polled from the queue the matching for the rewriting is encoded in the queue number already (7).}
	\vspace{-0.3cm}
	\label{fig:lb_vnf}
\end{figure}

As mentioned before, we use the matching capabilities of the NIC to increase the throughput of HNLB.
Figure \ref{fig:lb_vnf} shows the implementation of HNLB, supporting hardware table offloading. Accordingly, the load balancing steps can be presented as follows: 
\begin{itemize}
	\item[(1)] Packets from new or unseen connections do not match any rule in the Flow Director table. Therefore, they are forwarded to the default queue of the NIC, i.e., Queue 0.
	\item[(2)] The developed software in HNLB polls all queues in a round robin manner for packets. As packets are processed in bursts, it cannot be guaranteed that the second packet of one connection is already matched by the hardware table. Therefore, packets' headers from the default queue are hashed and checked against the software connection table. 
	\item[(3)] If the packets do not match any entry in the connection table, a DIP is chosen from the VIP table as usual.
	\item[(4)] The resulting mapping is installed in the software connection table (4a) and the hardware table (4b). 
	\item[(5)] The packet is rewritten according to the selected VIP-to-DIP mapping and put into the output buffer of the NIC, where it is forwarded.
	\item[(6)] Subsequent packets are matching the new rule in the hardware table and put into the queue that encodes the DIP. 
	\item[(7)] The header of these packets are then rewritten in software accordingly and are put to the output buffer of the NIC to forward them.
\end{itemize}

We based our implementation on DPDK~\cite{2014data}, which is a framework enabling high performance packet processing implementations.


For comparison, we also developed a SLB. To perform this, we use the implementation from Figure \ref{fig:lb_vnf} without step 4b.
This implementation does not use any hardware offloading, since it is realized only in software.
As a result, no hardware table is used, and only the default queue (Queue 0) is polled.


\section{Assessing utilization}
\label{sec:ass}
In this section, we introduce a utilization metric and a novel algorithm to determine it.

One of the techniques that DPDK uses to increase the packet throughput is the change from an interrupt-based packet retrieval to a polling-based packet retrieval. 
For instance, if a packet arrives on the NIC, the CPU is interrupted and the packet is then copied and processed by the OS. 
However, DPDK does not use interrupts, instead, it checks for packets at the NIC, processes these packets and then checks again for packets in an infinite loop. 
As a consequence, the conventional CPU utilization metric does not reveal the load of the system as the CPU is always fully utilized by the loop. 
In fact, our proposed metric can be useful to monitor the system and also to scale up the load-balancer in high-load conditions.

To overcome the challenges the metric shall fulfill the following conditions for a constant number of concurrent connections and packet sizes:
\begin{itemize}
	\item[(i)] The utilization shall reach \unit[100]{\%}, when the maximum possible packet rate is being processed and there is no packet loss.
	\item[(ii)] The utilization with no traffic (packet rate of 0 pps) shall be 0.
	\item[(iii)] Otherwise (when conditions (i) and (ii) are not met), it should be linear with the packet rate.
\end{itemize}


To form our metric, we adapt the algorithm presented in \cite{Sieber:2017:TOA:3155921.3158429}.
The gathering of the metrics works as follows:

The cycles counter is read before every iteration of the loop, the cycles spent in the last iteration are computed in line 3 and added up to the $REF$ counter.
If packets are processed, these CPU cycles that are used for processing packets are summed up in line 13 and are called $OPS$.
If we output $REF$ and $OPS$, we can compute the utilization in some time interval as:
\[util=\frac{OPS}{REF}\]
Therefore, if packets are processed in every iteration, we have $util=100\%$. 
Otherwise, with busy waiting iterations present, $util$ would be smaller than $100\%$.
However, this simple definition does not take into account that packets are processed in bursts:
If every iteration of the loop would process exactly one packet, the resulting utilization is 100\%, although the system can cope with higher rates.
In fact,we observe these effects in both cases. 
For instance, for the offloading case, with more used queues the queues are empty less often when polled. Therefore $util$ overestimates the utilization.
As a result, $util$ can fulfill conditions (i) and (ii), but fails for condition (iii).

\begin{algorithm2e}[t]	
	\caption{Receiving-Loop} 
	\label{alg1}
	\SetAlgoLined
	\BlankLine
	\While{true}{
		cpu\_cycles\_before = get\_cycles()\;
		REF+= (cpu\_cycles\_before - cpu\_cycles\_last)\;
		cpu\_cycles\_last = cpu\_cycles\_before\;
		number\_rx\_packets = recieve\_function()\;
		\If{number\_rx\_packets $>$ 0}{
			$n_p$ += number\_rx\_packets\;
			$n_b$ += 1\;
			$\dotsb$\\
			Packet processing\\
			$\dotsb$\\
			cpu\_cycles\_proc = get\_cycles()\;
			OPS += (cpu\_cycles\_proc - cpu\_cycles\_before)\;
		}
	}
\end{algorithm2e}

In order to resolve this, we also count the number of processed bursts $n_b$ and the number of processed packets $n_p$ in the lines 7 and 8.
This leads to an improved utilization metric presented as below:
\[util_+=util\cdot(1+\frac{n_p}{n_b\cdot B})/2\]
where $B$ is the maximum feasible burst size.
Essentially, $util_+$ weights $util$ with the mean burst utilization.

From our experiments, we can see that $B=32$ in the SLB case, which is exactly the maximum number of packets we read in one loop iteration from a queue.
For more than one queue, i.e. HNLB, this value can never be reached without losing packets. Therefore, we set $B=16$ and adjust the $util_+$ equation slightly:
\[util_+=util\cdot min(1,(1+\frac{n_p}{n_b\cdot B})/2)\]

%
%



\section{Performance Evaluation}
\label{sec:eval}
In this section, we compare HNLB with the SLB approach as described in Section \ref{sec:Arch}.
We focus on evaluating performance by measuring two metrics: \textit{i)} maximum throughput (without loss), and \textit{ii)} our novel utilization metric defined in Section \ref{sec:ass}.
Maximum throughput directly relates to the resource consumption of the load balancer, as more servers are needed, if less throughput per core is achievable.
The utilization on the other hand, gives more insights into the reasons of the throughput gain.
Additionally, it is necessary for scaling as packet loss should be avoided in real setups:
The utilization can be used as input metric for a scaling solution.
Service instances can be scaled up/down to improve resource efficiency depending on the utilization.

All results only refer to a single physical CPU core without exploiting parallelisms.
All measurements were repeated 30 times and confidence intervals were derived.
Since all results are very stable, the confidence intervals are not visible and, consequently, not shown.
The results are shown with respect to the sending rate of the traffic generator in million packets per second (mpps).


Our used testbed includes a Desktop PC which uses DPDK-Pktgen \cite{dpdk-pkt} for the load generation and a server equipped with 2x AMD EPYC 7301 x86 processor. 
Each of the processors has 16 Cores and 64MB L3 Cache.
The turbo feature was disabled as it mainly increases the performance of single-threaded applications, which is not a realistic use case on such a server.
The Simultaneous Multi-threading (SMT) support was disabled as well to avoid possible contention and interference effects, which are inherent of this technology. 
Moreover, both devices are equipped with Intel X550 Dual-Port 10G Ethernet interface cards.
We note that only one port of both NICs is used for the measurements.

\subsection{Maximum Throughput}

Let us first explore the maximum achievable throughput in terms of packet rate.
To perform this experiment, we use a single flow and a single DIP.

\begin{figure}[tb]
	\centering
	\includegraphics[width=0.9\columnwidth]{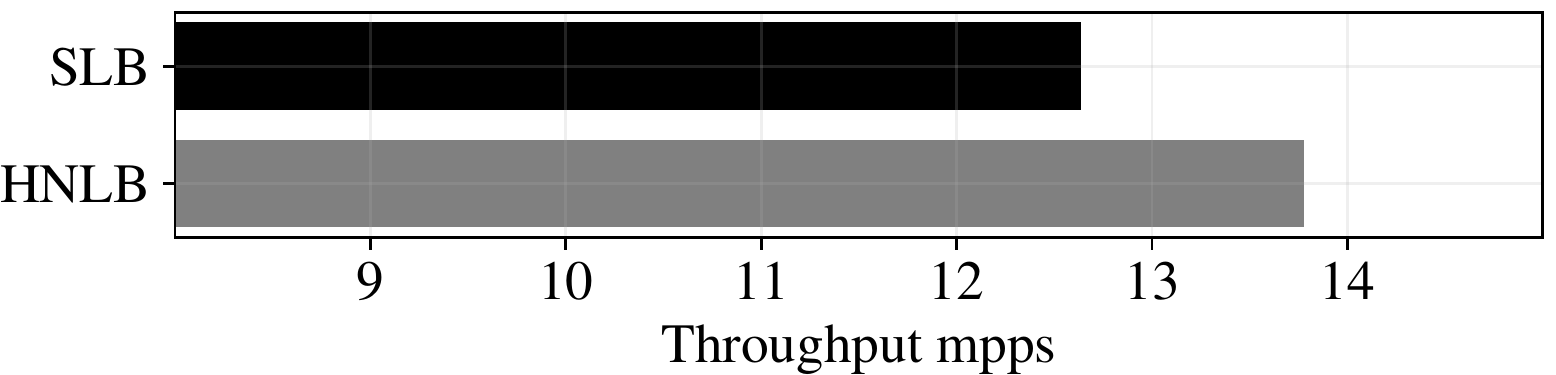}
	\vspace{-0.3cm}
	\caption{Best Case Packet Throughput for a paket size of 64 Byte, a single flow and a single DIP.}
	\label{fig:ThBC}
\end{figure}

As it is depicted in Figure \ref{fig:ThBC}, both solutions can achieve more than 12 mpps of packet rate.
In detail HNLB can achieve packet rates up to more than 13 mpps until it shows some loss (not shown in the figure).
However, in SLB case, packet loss starts when the packet rate reaches around 12 mpps. 
This resembles the overhead of reading and hashing the header in the SLB case, that is not necessary in the HNLB implementation.

The number of queues in HNLB is equal to the number of DIPs per core, i.e.only one queue is used to explore the best case.
For the following experiments, we use 10 queues.

\begin{figure}[tb]
	\centering
	\includegraphics[trim=0.5cm 0.7cm 0.5cm 0.7cm,clip,width=0.9\columnwidth]{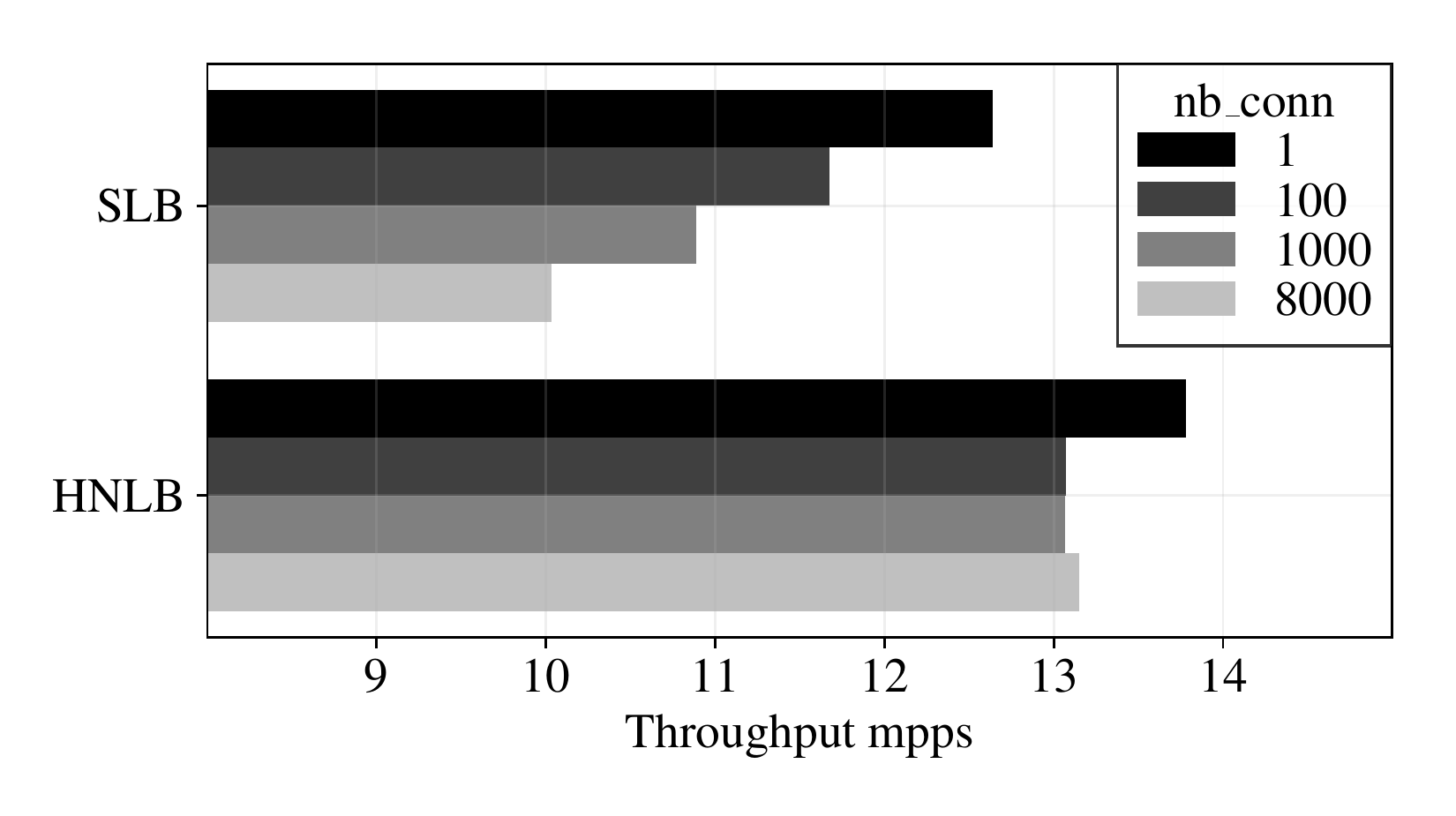}
	\vspace{-0.3cm}
	\caption{Maximum Throughput for a packet size of 64 Byte and different number of concurrent connections.}
	\label{fig:Loss_multiflow}
	\vspace{-0.3cm}
\end{figure}

In our next experiment, we investigate the throughput for different number of concurrent connections (nb\_conn).
In this case, we use only packet size equal to 64 Bytes. 
This experiment is depicted in Figure \ref{fig:Loss_multiflow}.
With an increased number of open connections, the application has to access a bigger table.
This decreases the locality of reference of the application, such that the processor cannot store the table in the fast on-chip caches anymore; hence, it has to access the main memory more often.
As a result, as is expected, higher number of connections reduces the maximum rate for both approaches.
For HNLB the throughput decreases from 1 to 100 connections, is constant for 100 to 1000 concurrent connections and even increases very little for 8000 connections.
The SLB approach throughput constantly decreases significantly with increasing number of connections. 
As a result, the performance gap between both approaches increases with more connections. For example, for 8000 parallel connections, HNLB can serve a 50\% higher packet rate than the SLB on one core.


\begin{figure}[tb]
	\centering
	\includegraphics[trim=0.5cm 0.7cm 0.45cm 0.7cm,clip,width=0.9\columnwidth]{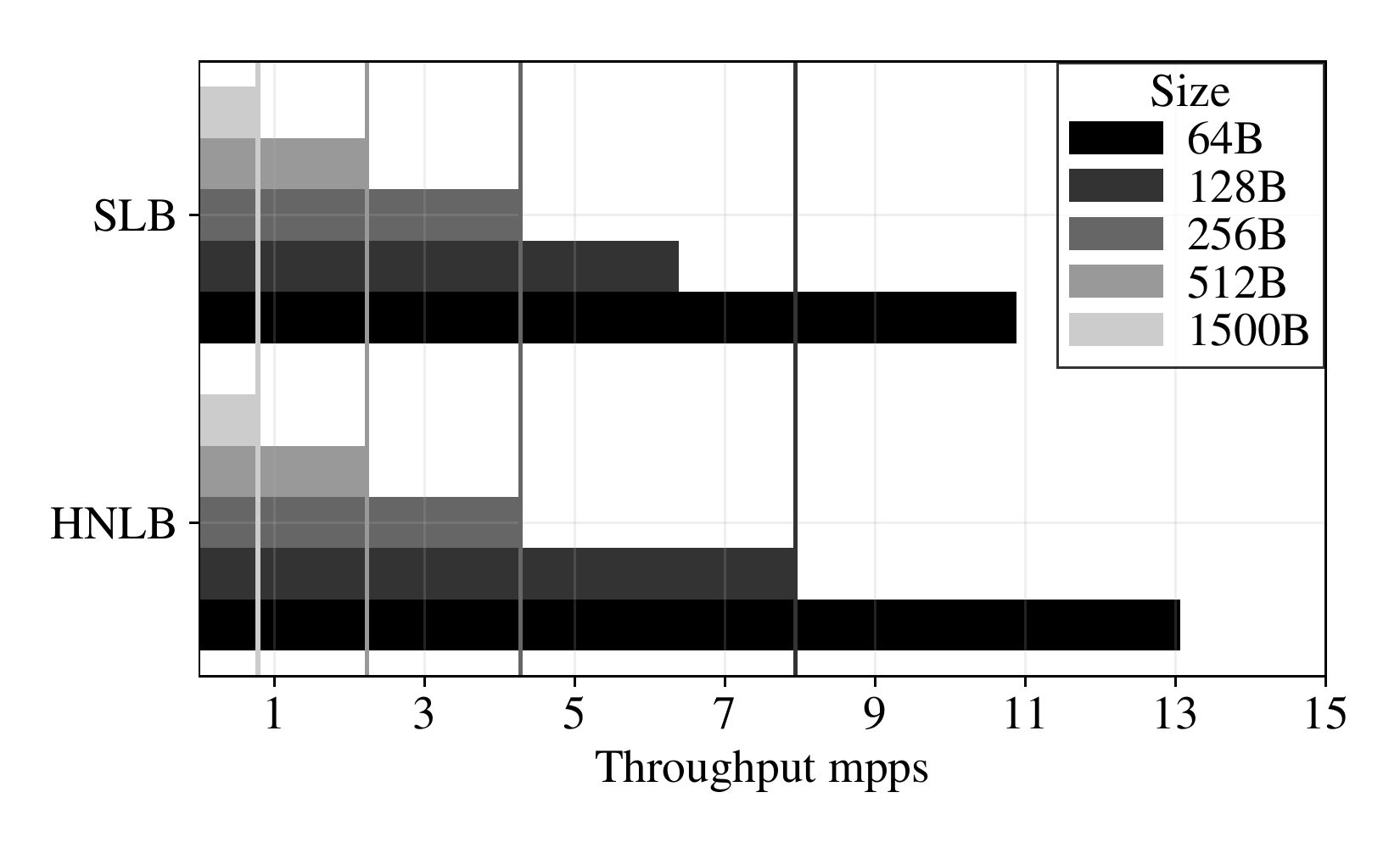}
	\vspace{-0.3cm}
	\caption{Maximum throughput for different packet sizes. Vertical lines represent the maximum offered rate of the traffic generator.}
	\label{fig:Loss_multisizes}
	\vspace{-0.3cm}
\end{figure}

We also evaluate the impact of packet sizes on the throughput in both cases.
Figure \ref{fig:Loss_multisizes} shows the maximum throughput of both approaches, where the vertical lines indicate the maximum rate of the traffic generator.
This maximum is not shown for 64 Byte packets as it is out of the scale.
In order to get stable results the maximum rate of the generator is a little below the NIC limit.

The results show that both solutions can reach the maximum rate of the traffic generator for packet sizes larger than 128 Bytes without any loss.
For a packet size of 128 Bytes the gain of HNLB is still at least 43\%, as HNLB can serve the maximum offered rate while SLB is already limited beforehand. 
The packet rate for larger sizes is limited by the rate of the traffic generator. 
We expect that the gain of HNLB will be similar for larger packet sizes.
This is supported by the utilization results in the next section.

\subsection{Utilization}
\label{sec:util_eval}


%
%

\begin{figure}[tb]
	\centering
	\includegraphics[width=0.9\columnwidth]{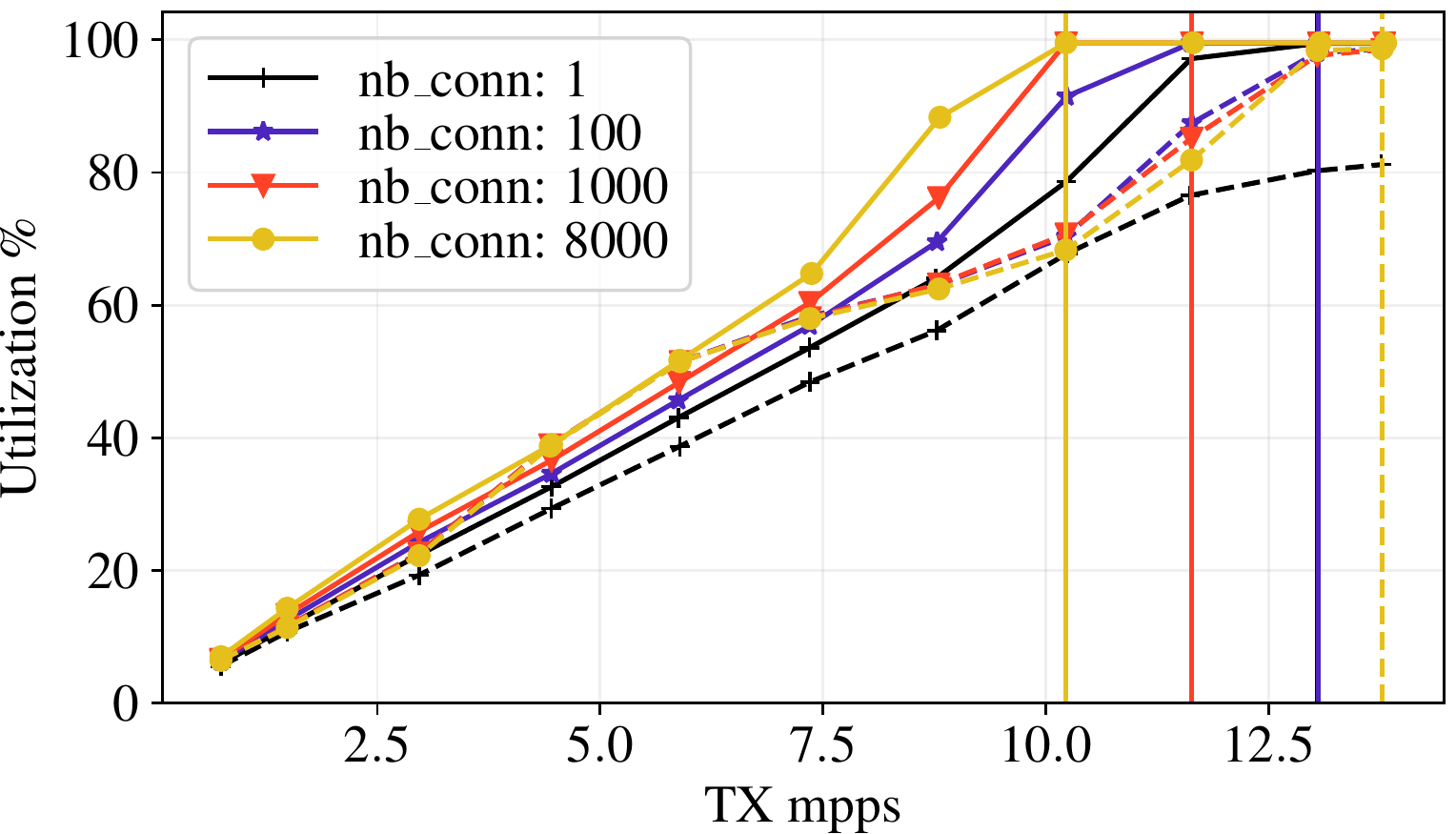}
	\vspace{-0.3cm}
	\caption{$util_+$ for different rates, different number of concurrent connections, a packet size of 64 Byte and 10 Queues. Vertical lines show the measurement values where the first loss occurs. HNLB is marked with dashed lines. SLB is shown with solid lines.}
	\vspace{-0.1cm}
	\label{fig:Util2_connections}
\end{figure}

Figure \ref{fig:Util2_connections} shows $util_+$ for both approaches and different number of concurrent connections.
The vertical lines show the packet rate where the first loss occurs.
From the Figure, it can be derived that HNLB outperforms the SLB approach, especially if a higher number of concurrent connections are served.
For instance, for 1000 connections and a packet rate of 10 mpps, the SLB has already a utilization of $100\%$, while HNLB has only a utilization of $70\%$. 
Further, we can argue that $util_+$ is a good metric to measure the utilization.
It gives a good indication of the load and how much more traffic can possibly be served:
For 1000 connections $util_+$ reaches 100\% with 10 mpps in the SLB case we and 13 mpps with HNLB.
For the SLB case we have already loss with 10 mpps, for HNLB loss starts with the next measurement value.
This shows that $util_+$ fulfills conditions (ii) and shows a close to linear behavior (condition(iii)) as well.
Further, $util_+$  fulfills condition (i) per definition.
$util_+$ with one connection on HNLB is always lower than 100\%, in reality this is not an issue as load balancing is not possible for one connection.

\begin{figure}[tb]
	\centering
	\includegraphics[width=0.9\columnwidth]{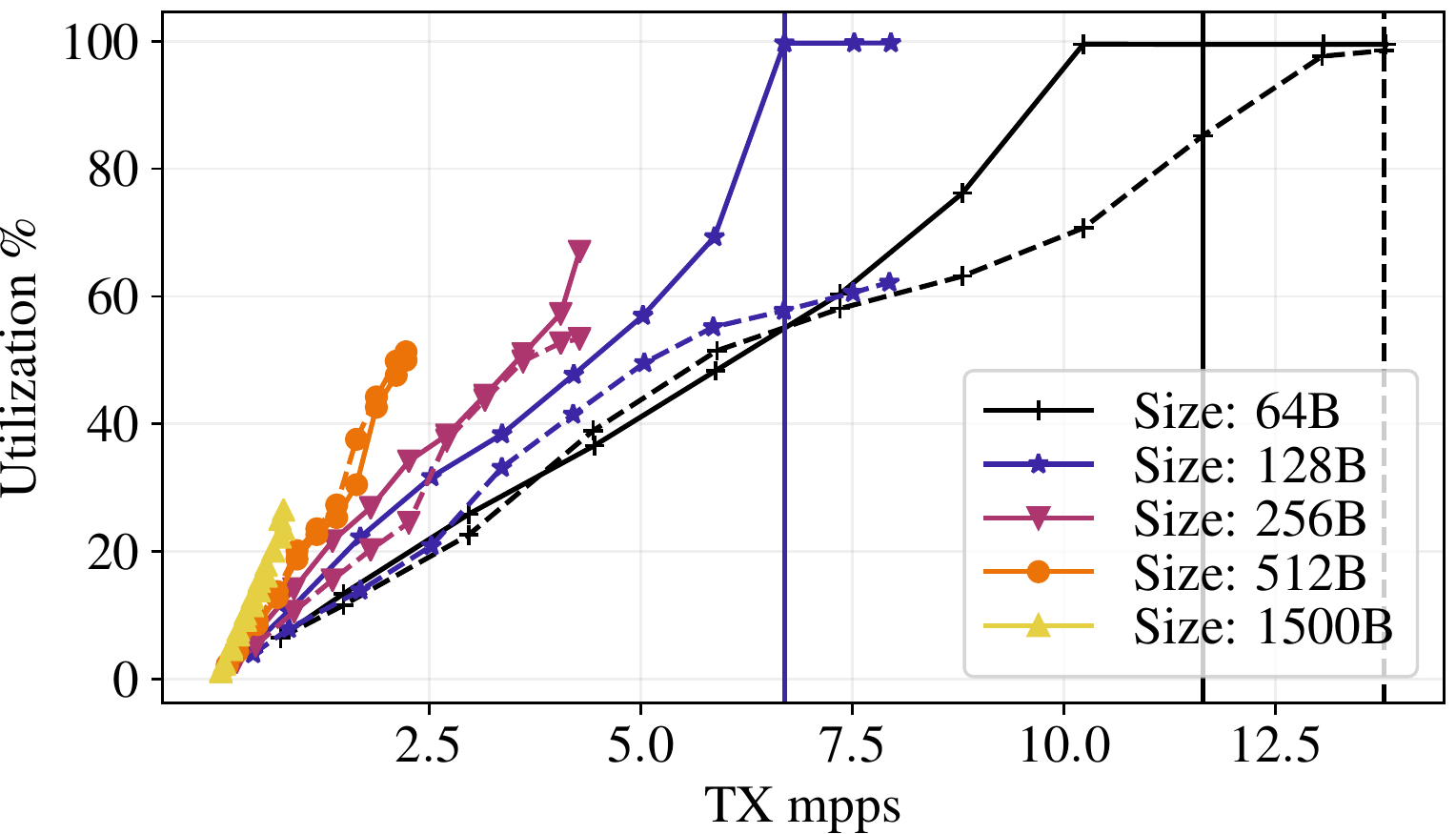}
	\vspace{-0.3cm}
	\caption{$util_+$ for different rates, different packet sizes, 1000 concurrent connections and 10 Queues. Vertical lines show the measurement values where the first loss occurs. HNLB is marked with dashed lines. SLB is shown with solid lines.}
	\label{fig:Util2_sizes}
	\vspace{-0.3cm}
\end{figure}

As our final experiment, Figure \ref{fig:Util2_sizes} shows the utilization for larger packet sizes. 
Since there is no loss in both cases for packets larger than 128 Bytes, we can observe the scalability with respect to larger packets using $util_+$ only.
We know that packets have to be copied to the main memory for processing and forwarding. Thus, larger packet sizes cause a higher utilization.
Nevertheless, we can clearly observe that HNLB does scale better for high packet rates. However, for low rates, both approaches have a comparable utilization or sometimes the SLB has even a little lower utilization.
This can be explained by the large number of small bursts that have to be processed by HNLB in these cases.
Even though we do not reach the limit of both approaches for larger packets, we still can observe that $util_+$ behaves similarly in regions we can cover with our setup.
Consequently a gain that is similar to the gain using small packets can be expected as well for high packet rates with large packets.

\section{Discussion}
\label{sec:dissc}

\begin{figure}[t]
	\centering
	\includegraphics[width=0.9\columnwidth]{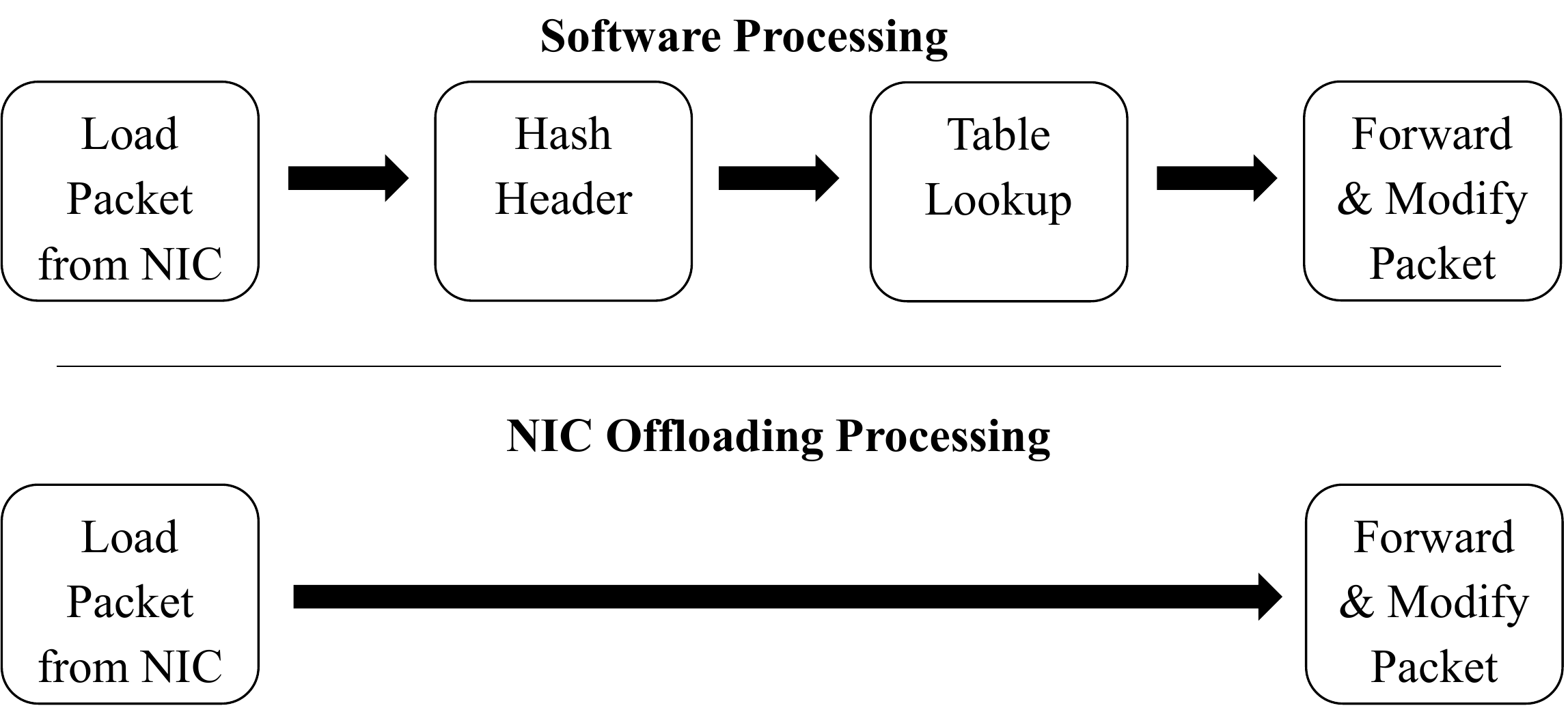}
	\vspace{-0.1cm}
	\caption{Software packet processing without Offloading and the matching capabilities. The flow assumes that a DIP table entry exists which holds for all packets of a connection except the first one.}
	\vspace{-0.5cm}
	\label{fig:pipelines}
\end{figure}

Section \ref{sec:Arch} discusses the approaches from an implementation point of view.
In this way, the reasons for the potentials of our solution and the gains visible in the results do not become clear immediately.
Therefore, in this section we highlight the processing steps that contribute most to the CPU load and can be improved by hardware offloading.
One important thing to note is, most of the packets match the DIP table directly, as only for the first packet of one connection the DIP can be chosen freely from the VIP table.
In our implementation, the first packet must always be processed in software; therefore, no gains are possible with NIC offloading in this case.
Figure \ref{fig:pipelines} shows the potential gains of the NIC Offloading approach, compared to the software approach with this restriction.
The goal of our approach is to utilize Flow Director matching capabilities in order to increase the throughput of the stateful load balancer; and in turn, reduce the packet loss.
Since Flow Director is not able to forward or modify packets, these two steps have to be done in the Offloading and the Software Processing case.
Nevertheless, we can utilize Flow Director for matching and lookup.
In software, this is implemented using a hash over the 5-tuple and a table lookup.
Both operations are considered to be $O(1)$. 
However, the table lookup can be comparatively costly, if modern CPU architectures are considered \cite{Sieber:2017:TOA:3155921.3158429}.
These architectures are designed for heavy computation tasks mainly.
One main point is the CPU cache architecture:
As long as all the data that is needed for computation is fitting in the caches, the performance would be high.
On the other hand, if the cache size is exceeded e.g. for slightly larger lookup tables in our case, the performance would decrease drastically.
Consequently, we can observe a performance drop of the SLB implementation if we use larger tables.
Additionally, the gain of NIC Offloading with small tables is smaller, as we can only save the computation cost of the hashing.

\section{Related Work} \label{sec:related_work}
In this section we give an overview on related work in the fields of Load Balancers, Hybrid Load Balancers and also NIC-Offloading.


\subsection{Hardware Load Balancers}
Traditionally, server load balancing has been achieved by dedicated hardware load balancers \cite{a10,f5,lb}. However, they have a number of issues. For instance, they are costly, complex to deploy, and more importantly, difficult to scale up/down based on the demand load \cite{patel2013ananta}. 

\subsection{Software Load Balancers}
With the introduction of network softwarization and Network Function Virtualization (NFV), many network functions have started to being deployed as software instances on commodity servers \cite{mijumbi2016network}. Therefore, many software load balancers have received attention \cite{patel2013ananta,eisenbud2016maglev}. Ananta \cite{patel2013ananta} is a software load balancer that uses commodity hardware to deploy software Muxes (SMux) instances to build a distributed data plane for traffic splitting and encapsulation. Ananta uses a central controller that utilizes SMux instances to map VIPs to DIPs by using ECMP. 

Google's Maglev \cite{eisenbud2016maglev} is another state-of-the-art software load balancer that can be deployed on commodity Linux servers. Since Maglev is deployed in software, the load balancing service capacity can be increased/decreased by simply adding/removing servers. Maglev utilizes consistent-hashing to load balance the traffic across service instances, regardless of which load balancer instance is forwarding the packet. In this way, it forwards the packets belonging to same connection to the same service instance. Benefiting from its distributed architecture, consistent hashing, and connection tracking, it can reduce the impacts of sudden failures and changes in the network.

\subsection{Hardware and (Hybrid) Load Balancers}

Despite low cost and high availability/flexibility of software load balancers, they suffer from high latency and low throughput \cite{gandhi2015duet}. These issues can be tackled by developing load balancers utilizing hardware. One of the candidate hardwares to enhance load balancing, is ASICs in network switches. Comparing to software load balancers, hardware (network switch)-based load balancers can process the same amount of traffic with about two orders of magnitude lower costs \cite{Miao2017}. 

Using these features, authors in \cite{gandhi2015duet} propose Duet, a hybrid load balancer that proposes to move the load balancing function to existing hardware network switches (at no extra cost). In fact, network switches perform traffic splitting (using ECMP) and packet encapsulation tasks of load balancing. Notably, VIP-to-DIP table is stored in switching ASICs, which can contribute to reduce the latency to several microseconds. In addition to hardware load balancing, they developed a small software load balancer to act as a backstop, and to provide high availability and flexibility.

Similar to Duet, SilkRoad \cite{Miao2017} also leverages features of programmable ASICs to build a load balancer for data center networks. In contrast to Duet \cite{gandhi2015duet} that stores (not per-connection) VIP-to-DIP mappings in switching ASICs, SilkRoad utilizes SRAM in ASICs and stores per-connection states at ASICs. In this way, in addition to providing high throughput and low latency, SilkRoad ensures per-connection consistency during DIP pool changes. However SilkRoad requires specialized switching hardware usually not available in DataCenters.

Conversely, there are works that use Software Defined Networking (SDN) to provide load balancing service \cite{handigol2009plug,wang2011openflow,kang2015efficient}. In these approaches, the load of service instances and network are being monitor by monitoring entity(es). Then, the SDN controller installs the corresponding forwarding rules on network switches to direct flows to proper service instances.

\subsection{NIC Offloading}
In addition to switching ASICs, NIC offloading mechanism can be used for deploying hybrid hardware-software load balancers and other stateful network functions. As an example for using NIC offloading, authors in \cite{weinsberg2006putting} offloaded the firewall logic to a NIC. They utilized 5-tuple filtering in the NIC and showed that a offloading firewall to NIC can improve both CPU utilization as well as packet throughput.

Further, authors in \cite{shi2016ndn} proposed NDN-NIC, a network interface card for performing name-based filtering on the NIC. 
In this approach, names are maintained in a bloom filter, which is used as a reference for filtering the incoming packets on NIC. 
Authors showed that this approach is able to reduce CPU overhead and energy consumption. 

In a recent work, Microsoft researchers [22] presented a NIC Offloading mechanism which uses custom NICs with a built-in FPGA. 
Their solution is used in Microsoft’s Azure cloud to reduce the CPU load caused by networking. 
Moreover, their approach uses an offloading method that handles the first packet in software. 
In contrast to our work, a more feature-rich FPGA is utilized,  while we are only using existing capabilities of the NIC.

However, to the best of our knowledge, we are the first work that present a hybrid hardware-software load balancer in which NIC offloading (Intel Flow Director \cite{flowdirector2014intel}) is utilized for offloading matching to hardware.

%

\section{Conclusion and Future Work}\label{sec:conc}

In this work, we present HNLB, a high performance hybrid hardware-software load balancer. 
We used Flow Director technology, available in modern high-speed server NICs, to perform NIC offloading.
We showed that by utilizing hardware matching capabilities, it is possible to increase the throughput of stateful network functions.
Especially, if a higher number of concurrent connections have to be processed, the throughput can be increased quite significantly.
For example, for 8000 concurrent connections, throughput can be increased by $50\%$ compared to the software-only case.
This is due to the fact that the software only solution has to maintain a table in software and directly reflects the effects of the CPU cache overload.
On the other hand, the employed NIC limits the number of concurrent connections to 8000.

Additionally, in this work, we assess the utilization of both solutions.
It is necessary to estimate the utilization of packet processing systems in order to be aware of overload situation in advance, i.e. before any packets are lost.
Further, to increase the performance, we use busy polling, which renders the CPU utilization useless, as it is always $100\%$.
As a solution, we proposed a metric that combines burst size measurements and the waiting cycles during busy waitings.


As future work, we plan to extend HNLB with a dynamic offloading system, in order to support a higher number of concurrent connections.
In reality, only a subset of the connections causes a large share of the rate in packet switched systems.
Therefore, it is sufficient to identify these connections and offload them to hardware, accordingly.
Furthermore, we plan to develop an architecture that supports an arbitrary number of DIPs by using a multistage architecture.

\section*{Acknowledgements}

This work has been partially funded by the German Research Foundation (DFG) under the grant number KE 1863/8-1 and under the Celtic-Plus subproject SEcure Networking for a DATa center cloud in Europe (SENDATE)-PLANETS (Project ID 16KIS0261, Celtic-Plus project ID C2015/3-1) funded by the German Federal Ministry of Education and Research (BMBF).
\bibliographystyle{IEEEtran}

\bibliography{bibliography}

%
%
%


\end{document}